\documentclass[pdflatex,sn-mathphys-num]{sn-jnl}% Math and Physical Sciences Numbered Reference Style

\usepackage{graphicx}%
\usepackage{multirow}%
\usepackage{amsmath,amssymb,amsfonts}%
\usepackage{amsthm}%
\usepackage{mathrsfs}%
\usepackage[title]{appendix}%
\usepackage{xcolor}%
\usepackage{textcomp}%
\usepackage{manyfoot}%
\usepackage{booktabs}%
\usepackage{algorithm}%
\usepackage{algorithmicx}%
\usepackage{algpseudocode}%
\usepackage{listings}%

\hypersetup{
  colorlinks=true,
  linkcolor=blue!50!black,
  citecolor=blue!50!black,
  urlcolor=blue!50!black
}

\newcommand{\dune}{DUNE}
\newcommand{\sm}{Standard Model}
\newcommand{\nd}{Near Detector}
\newcommand{\cpv}{CP violation}
\newcommand{\pmns}{PMNS}
\newcommand{\ckm}{CKM}

\begin{document}

\title{Flavor as an Incomplete Structure: \\ Conceptual Questions and the Role of \dune}

\author*[1,2]{\fnm{Claudio Silverio} \sur{Montanari}\email{cmontana@fnal.gov}}

\affil*[1]{\orgname{Fermi National Accelerator Laboratory}, \orgaddress{\city{Batavia}, \state{Illinois} \postcode{60510}, \country{USA}}}

\affil[2]{On leave of absence from INFN Sezione di Pavia, Pavia, Italy}

\abstract{Flavor remains one of the most successful yet least understood structures of the \sm.
The discovery of the Higgs boson completed the electroweak account of mass generation,
but did not explain the origin of fermion families, mass hierarchies, or mixing patterns.
In this sense, flavor can be regarded as an empirically successful but conceptually incomplete structure.
Neutrinos occupy a particularly sensitive place within this problem: their masses are tiny,
their mixing is large, and their mass-generation mechanism may differ from that of charged fermions.
In this article, we discuss flavor as an open conceptual problem and argue that \dune,
as a phased program spanning precision oscillation measurements and sensitivity to BSM and
dark-sector phenomena, provides a powerful framework for testing the self-consistency and
possible limits of the present three-flavor description. 
In particular, the complementarity between the long-baseline program and the Phase I 
near-detector complex, together with the DUNE-PRISM strategy for controlling interaction-model 
systematics and enabling data-driven near-to-far predictions, makes \dune\ especially well-suited 
to search for small, correlated departures from the minimal flavor framework.}


\maketitle

% -------------------------------------------------
\section{Introduction} \label{sec1}
% -------------------------------------------------

The gauge sector of the \sm\ provides a remarkably economical and predictive description
of fundamental interactions, yet its flavor sector remains largely empirical.
Fermion masses and mixings are encoded in Yukawa couplings, whose values span many orders
of magnitude and exhibit no known organizing principle.
The discovery of the Higgs boson at the Large Hadron Collider (LHC) established
the mechanism by which fermion masses arise through electroweak symmetry breaking
and Yukawa interactions~\cite{bib1,bib2,bib2_1,bib3,bib4}.
In this sense, flavor is not an auxiliary feature of the theory:
it is the structure through which fermion mass becomes physically manifest.
At the same time, the Higgs mechanism does not explain why flavor takes
its observed replicated and hierarchical form.
The flavor sector therefore remains an empirically successful,
yet conceptually incomplete, component of the \sm.

This incompleteness is particularly evident in the existence of three fermion families
with identical gauge quantum numbers but distinct masses and mixings.
While the quark sector is described by the \ckm\ matrix with small mixing angles,
the lepton sector is characterized by the \pmns\ matrix, which exhibits large mixing
and potentially large, but not yet established, \cpv\ effects~\cite{bib6,bib7,bib8}.
Neutrino oscillation experiments have firmly established that neutrinos are massive
and mix across flavors~\cite{bib9,bib10,bib11}. This fact alone requires an extension 
of the minimal \sm\ to accommodate non-vanishing neutrino masses.
However, the very nature of this extension is fundamentally undetermined, since key aspects of the neutrino sector are still open:
the absolute mass scale, the ordering of mass eigenstates,
the presence and size of leptonic \cpv, and the Dirac or Majorana nature of neutrinos.
More fundamentally, it is not known whether neutrino masses arise from the same
Yukawa mechanism as charged fermions or from qualitatively different dynamics,
such as seesaw constructions or other beyond-the-\sm\ (BSM) scenarios~\cite{bib12,bib13,bib14}.

Recent reviews of flavor model building emphasize that the field has developed a broad 
toolkit---including non-Abelian discrete symmetries, continuous horizontal symmetries, 
texture-based constructions, modular-invariant frameworks, and seesaw-motivated realizations 
of neutrino mass---that has clarified how observed hierarchies 
and mixing patterns may arise and, in selected cases, yields useful correlations or partial 
predictions~\cite{bib15,bib16}. At the same time, these reviews also 
make clear that realistic agreement with data often requires additional assumptions, such as 
non-unique charge or group assignments, extra spurions or fields, or threshold structure beyond 
the minimal setup~\cite{bib15,bib16}. In many cases, the resulting 
low-energy consequences are not unique and do not furnish distinctive smoking-gun signatures 
at accessible energies~\cite{bib15,bib16}. From this perspective, 
the model-building literature is highly valuable and instructive, but it also reinforces the need 
for precision experimental probes capable of testing whether the effective flavor framework 
is fully self-consistent. In the present work, this complementary viewpoint is adopted with a focus 
on neutrino measurements as an experimentally accessible arena in which such limits can be explored.

The question for neutrino experiments is therefore not whether they can solve the full flavor
problem of the \sm\ directly, but whether the minimal three-flavor description of
neutrino mixing and interaction remains internally consistent when tested at high
precision. In this sense, neutrino measurements provide a window onto the broader
flavor problem without exhausting it.

From this perspective, neutrinos occupy a privileged position within the flavor problem.
Their small masses and large mixing angles suggest that they may probe a different 
regime of flavor structure, one potentially sensitive to physics beyond the minimal three-flavor paradigm.
If the flavor sector is not fundamentally complete, deviations from the standard picture
may first appear as subtle inconsistencies rather than dramatic new states.
Examples include non-unitarity of the \pmns\ matrix, non-universal couplings,
or the presence of additional interactions that distinguish among flavors.
Such effects are often parametrized in terms of non-standard neutrino interactions (NSI),
which provide a general framework to describe deviations from the weak interaction structure
of the \sm~\cite{bib17,bib18,bib19,bib20}.
Similarly, a broad class of BSM scenarios---including sterile neutrinos, hidden sectors,
and flavor-dependent mediators---can give rise to small but measurable departures
from standard oscillation and interaction patterns~\cite{bib21,bib22}.

In this context, the experimental exploration of neutrino flavor becomes a probe not 
only of oscillation parameters, but also of the self-consistency of the neutrino-flavor 
sector. In this sense, it may also provide access to a deeper underlying flavor structure. 
The Deep Underground Neutrino Experiment (\dune) is particularly
well positioned to address this program within the neutrino sector~\cite{bib23,bib24,bib25}.
Its long-baseline component is designed to measure the neutrino mass ordering,
leptonic \cpv, and oscillation parameters with unprecedented precision.
At the same time, the \nd\ complex provides high-resolution measurements of neutrino interactions
in a controlled environment, enabling detailed studies of cross sections,
kinematics, and flavor-dependent effects~\cite{bib26}.

The complementarity between near and far detectors is essential.
While long-baseline measurements probe neutrino propagation over macroscopic distances, 
the \nd\ constrains the unoscillated beam and the interaction model near the source, and 
provides direct access to precision neutrino interaction measurements.
This dual capability allows \dune\ to disentangle effects arising from propagation,
interactions, and potential new physics.
In particular, the \nd\ offers a powerful platform to search for small deviations
from the minimal flavor framework, including non-standard interactions,
non-unitarity, and signals associated with hidden or dark sectors~\cite{bib27}.

The central thesis of this work is that flavor should be regarded as an incomplete structure,
whose full organization is not yet understood.
Rather than proposing a specific model, we adopt a conceptual perspective:
we argue that the empirical success of the current framework does not imply its completeness,
and that neutrino experiments, and \dune\ in particular, provide a powerful opportunity
to test this hypothesis in the experimentally accessible neutrino sector.
In this view, the role of \dune\ extends beyond precision measurement of known parameters;
it becomes an experimental program capable of testing the self-consistency and possible
limits of the three-flavor framework, while searching for small departures that may point
to a deeper underlying organization of flavor.

% -------------------------------------------------
\section{Flavor as an Unfinished Sector of the Standard Model} 
% -------------------------------------------------

The \sm\ provides a remarkably successful and internally consistent description of
fundamental interactions. Its gauge structure is highly constrained and predictive,
leading to a coherent understanding of electroweak and strong phenomena.
In contrast, the flavor sector is largely empirical: fermion masses and mixings
are encoded in Yukawa couplings whose values are not explained within the theory.

In the minimal renormalizable \sm, the flavor sector is specified by Yukawa matrices
for the charged fermions. After electroweak symmetry breaking, these matrices generate
charged-fermion masses, and in the quark sector the mismatch between the up- and
down-type mass bases gives rise to the \ckm\ matrix. The observed lepton mixing
described by the \pmns\ matrix enters only once the theory is extended, or treated
effectively, to account for neutrino mass. This is precisely why neutrinos are so
informative for the flavor problem: the lepton sector is not only phenomenologically
distinct from the quark sector, but already indicates that the minimal formulation
is incomplete. Both the \ckm\ and \pmns\ matrices are determined experimentally,
yet they exhibit markedly different patterns.

Despite its phenomenological success, this structure raises several fundamental questions:
\begin{itemize}
    \item Why do fermions appear in three families with identical gauge quantum numbers?
    \item What determines the hierarchy of Yukawa couplings spanning many orders of magnitude\footnote{This issue becomes even more striking if neutrino masses are also tied to electroweak symmetry breaking, whether through tiny Dirac Yukawa couplings or effective operators involving the Higgs field.}?
    \item Why do quarks exhibit small mixing while leptons exhibit large mixing?
    \item Is the observed flavor structure complete, or only an effective description?
\end{itemize}

These questions highlight a tension within the \sm: while gauge interactions are
organized by symmetry principles, flavor appears as a collection of parameters
with no evident underlying structure. In this sense, flavor is not merely unknown,
but structurally underdetermined.

This motivates viewing flavor as an \emph{unfinished sector} of the theory.
By this we mean that the flavor structure is fully operational---capable of describing
all observed phenomena---yet lacks a conceptual closure comparable to that of the gauge sector.
The parameters of flavor are measured, but their origin and organization remain unexplained.

An important aspect of this incompleteness is the replication of fermion families.
Each generation carries the same gauge charges but differs in mass and mixing,
suggesting that flavor labels may not correspond to fundamental quantum numbers
in the same sense as gauge charges. Instead, they may reflect a deeper level of
organization not captured by the current formulation.

From this perspective, the Yukawa sector can be interpreted as an effective
parameterization of a more fundamental structure. The success of this parameterization
does not imply its completeness: it may instead represent the low-energy manifestation
of dynamics that remain hidden at accessible scales.

This point of view suggests that deviations from the minimal flavor framework
need not appear as large or isolated signals. Rather, they may emerge as small,
coherent inconsistencies across different observables, or as patterns that resist
simple interpretation within the existing structure. Identifying such deviations
requires both precision and a framework capable of recognizing departures from
expected correlations.

In this context, neutrinos occupy a particularly important position.
Their masses are much smaller than those of charged fermions, and their mixing
pattern differs qualitatively from that of quarks. These features suggest that
the neutrino sector may probe aspects of flavor that are not visible in the
charged-fermion sector, and may therefore provide a sensitive test of whether
the present neutrino-flavor description is complete or only effective.

Neutrinos are therefore not uniquely easy probes of flavor, but uniquely sensitive ones.
Their coherent propagation over long baselines makes small departures from the
three-flavor picture observable in principle, provided that flux, detector, and
interaction uncertainties can be controlled with sufficient precision.
It is precisely this experimental requirement---the need to control beam flux,
interaction modeling, and detector response while preserving sensitivity to small,
correlated deviations---that gives the \dune\ near-detector complex its central physics role.

% -------------------------------------------------
\section{The Higgs Mechanism and the Physical Emergence of Flavor} 
% -------------------------------------------------

The Higgs mechanism provides the framework through which charged fermions acquire physical masses.
Electroweak symmetry breaking transforms Yukawa matrices into mass matrices; after diagonalization,
one obtains physical mass eigenstates and the observable charged-current mixing structure.
Most notably, because the model contains no right-handed neutrinos, no analogous Yukawa interaction
can be written for neutrinos, and they are massless in the minimal renormalizable \sm.

However, the Higgs mechanism does not explain:
\begin{itemize}
    \item the origin of Yukawa couplings,
    \item the replication of fermion families,
    \item the hierarchy of masses,
    \item the distinct mixing structures of quarks and leptons.
\end{itemize}

In this sense, the discovery of the Higgs boson~\cite{bib1,bib2,bib2_1,bib3,bib4}
did not resolve the flavor problem, but rather sharpened it: it made the structure observable
and operational, without explaining its origin.

This shift can also be viewed at a conceptual level. The structure of spacetime already
admits both null and timelike trajectories; however, it is the existence of massive
particles that renders rest frames operational for matter fields. In this sense,
electroweak symmetry breaking does not modify spacetime itself, but provides the
dynamical structure through which massive degrees of freedom become physically realized
for charged fermions and electroweak gauge bosons. Any deeper organizing principle of flavor 
must therefore explain not only why the Yukawa couplings take their observed values, but also 
why the Higgs field is the unique mediator through which flavor labels acquire physical consequences.

If one adopts this viewpoint, a unified renormalizable origin of fermion masses acquires
a particular apparent conceptual economy: a single structure would underlie not only mass generation,
but also the physical emergence of massive kinematics across the Standard Model. Neutrinos
are critical precisely because they test whether this economy survives beyond the charged-fermion
sector. If their masses arise from Dirac Yukawa couplings, the unified picture is preserved,
albeit with extraordinarily small couplings. If instead they arise effectively, through higher-dimensional 
operators or from heavier states integrated out at higher scales, then electroweak symmetry breaking 
still enters the low-energy mass term, but the minimal renormalizable flavor structure is no longer 
sufficient\footnote{For example, the unique dimension-5 Weinberg operator generates Majorana 
neutrino masses after electroweak symmetry breaking~\cite{bib5}.}. In that sense, neutrinos identify the 
point at which the apparent unity of mass generation is tested most sharply.

The experimentally accessible question is therefore not the ultraviolet origin of neutrino mass itself, but whether
low-energy observables remain fully consistent with the minimal three-flavor framework.

% -------------------------------------------------
\section{Neutrinos as a Precision Test of the Three-Flavor Framework} 
% -------------------------------------------------

Neutrinos provide an especially sensitive arena in which to test flavor structure.
Their tiny masses and large mixing angles make flavor conversion observable over
macroscopic distances, so that small departures from the standard three-flavor
description can appear as correlated distortions in oscillation probabilities,
interaction rates, and reconstructed kinematic distributions.

The experimental advantage is not simplicity, but controlled redundancy: multiple detector 
elements and multiple flux configurations enable internal cross-checks and data-driven constraints. 
Accelerator-based neutrino experiments face substantial uncertainties from flux
prediction, neutrino--nucleus interaction modeling, nuclear effects, and detector
response. What makes \dune\ powerful is that it is designed to control these
effects through the complementarity of its long-baseline measurement and its
near-detector complex~\cite{bib26}.

From this perspective, DUNE should be viewed as an integrated precision apparatus 
rather than as a far detector supplemented by an auxiliary near detector. Its reach comes 
from the combination of a high-power, predominantly muon-flavor wide-band beam, a massive 
liquid-argon far detector with high-resolution imaging and calorimetric reconstruction, and a 
near-detector complex with movable, magnetized, and complementary subsystems able to 
characterize the beam and neutrino interactions close to the source. The role of this architecture 
is to turn the near-to-far connection into an increasingly data-driven procedure, so that spectral 
variation, charge selection, and continuous beam monitoring can constrain the dominant flux, 
interaction, and detector-response uncertainties toward the few-percent regime required 
by DUNE's precision program. The concrete realization of this strategy in Phase~I---through 
ND-LAr+TMS, SAND, and the DUNE-PRISM program---is discussed in Sec.~VI.B.

Because the beam is operated in both neutrino and antineutrino mode and spans the energy 
range of the leading oscillation channels while retaining sensitivity to tau appearance in 
its high-energy tail, DUNE can test flavor structure through several complementary channels 
rather than through a single inclusive rate~\cite{bib23}.

From this perspective, neutrino physics is not merely an extension of flavor
studies, but a precision test of the domain of validity of the three-flavor
framework. Small, coherent deviations across oscillation probabilities, flavor-tagged 
event samples, and reconstructed energy spectra may provide the first evidence that the
present description is incomplete. This motivates a search strategy based not
only on large, isolated anomalies, but also on internally correlated departures
from the expectations of the standard framework.

% -------------------------------------------------
\section{Observed Anomalies, Present Constraints, and the Need for Complementary Tests} 
% -------------------------------------------------

If the flavor sector is not fundamentally complete, its limitations may not manifest
as large or isolated signals, but rather as small, correlated deviations from the
predictions of the minimal three-flavor framework. In this regime, the relevant
observables are those that probe consistency across different channels: mixing,
interaction rates, and kinematic distributions. The experimental challenge is
therefore not only to achieve high precision, but also to identify patterns of
deviation that cannot be accommodated within the existing structure. Such patterns
may reflect non-unitarity of the mixing matrix, departures from universality in
neutrino interactions, or the presence of additional degrees of freedom that couple
weakly to the Standard Model.

These considerations suggest that the search for new physics in the flavor sector
should be framed as a program that combines consistency tests with direct rare-process searches.
Deviations may appear as tensions between measurements performed in different regimes,
or as systematic distortions that persist across datasets. Identifying such effects
requires a coherent experimental program in which multiple observables are measured
with sufficient precision and interpreted within a unified framework.

Possible experimental manifestations include:
\begin{itemize}
    \item deviations from \pmns\ unitarity,
    \item non-standard neutrino interactions,
    \item flavor-dependent distortions of cross sections,
    \item anomalous neutral-current processes,
    \item signatures of hidden or dark-sector couplings.
\end{itemize}

In this regime, discovery may correspond to identifying coherent patterns of small deviations
rather than large, isolated signals.

\subsection{Observed anomalies and tensions}

Short-baseline anomalies in LSND~\cite{bib28}, MiniBooNE~\cite{bib29},
reactor~\cite{bib30}, and gallium~\cite{bib31} data have long
motivated extensions of the minimal three-flavor picture, especially light-sterile scenarios.
The present situation, however, is mixed and increasingly constrained.
MicroBooNE excludes a single-sterile (3+1) oscillation explanation of the LSND/MiniBooNE anomalies~\cite{bib31_1},
STEREO and updated reactor analyses weaken the sterile-neutrino interpretation of the reactor anomaly~\cite{bib32,bib33},
and KATRIN excludes substantial portions of the parameter space associated with gallium-motivated
interpretations while also challenging the Neutrino-4 claim~\cite{bib34}.
At the same time, the gallium anomaly itself remains unresolved: recent reviews note that it
persists at approximately the \(4\sigma\) level under conservative assumptions, but BEST saw no
distance dependence, and the simplest \(3+1\) interpretation remains in strong tension with other
\(\nu_e\) and \(\bar{\nu}_e\) disappearance data~\cite{bib36,bib35}.

This does not weaken the case for \dune; rather, it strengthens the need for a complementary
precision program in which oscillation effects, interaction modeling, and direct rare-process
signatures can be disentangled with the near-detector complex and the long-baseline measurement.

% -------------------------------------------------
\section{The Role of the \dune\ Program} 
% -------------------------------------------------

A recurring limitation of some accelerator-based low-energy-excess searches has been the reliance on inclusive 
excesses in ``electronlike'' event samples, often with comparatively limited power to separate 
electrons from photons and limited access to the full hadronic recoil system. By contrast, liquid-argon 
time projection chambers provide finely segmented 3D imaging and calorimetry, so that an interaction is 
reconstructed not only as a rate in a signal window, but as a topology with an identified vertex, tracks, 
showers, local $dE/dx$, and visible hadronic activity~\cite{bib24,bib37}. In particular, 
the measurement of $dE/dx$ in the first centimeters of an electromagnetic shower, together with 
conversion-distance and vertex information, provides powerful electron/photon separation, while 
contained protons and other low-energy hadrons can be reconstructed and used as part of the event characterization~\cite{bib39,bib38}. Neutrino energy 
can then be estimated calorimetrically for electromagnetic activity and from range or calorimetry for 
contained tracks, yielding a far richer description of the visible final state than in earlier rate-driven 
searches~\cite{bib38,bib37}. MicroBooNE has already demonstrated 
the value of this approach in low-energy-excess studies by analyzing exclusive, semi-inclusive, and 
fully inclusive topologies with different hadronic final states, and by using topological and calorimetric 
observables to reject $\pi^0$ and single-photon backgrounds~\cite{bib38,bib39}. 
In DUNE, the same LArTPC capabilities at the near and far detectors make it possible to compare 
data and predictions in a multidimensional space of observables---including shower-start $dE/dx$, 
lepton angle, proton content, visible hadronic energy, and related kinematic correlations---rather 
than through a rate excess alone~\cite{bib24,bib27}. The experimental gain 
is therefore not simply higher statistics, but a more fully constrained event-level characterization 
that improves the separation of Standard-Model backgrounds, detector effects, and genuinely 
BSM contributions.

The \dune\ program~\cite{bib23,bib24,bib25} provides 
a uniquely powerful experimental framework
in which the structure of neutrino flavor can be probed across multiple
scales and observables. Its design combines a high-intensity neutrino beam,
a precision near-detector (\nd) complex, and a long-baseline far detector,
allowing for a coherent exploration of both neutrino propagation and
interaction dynamics.

\subsection{Long-baseline Measurements}

The long-baseline component of \dune\ is optimized to measure the fundamental
parameters governing neutrino oscillations. By measuring $\nu_e$ and $\overline{\nu}_e$ appearance and 
$\nu_\mu$ and $\overline{\nu}_\mu$ disappearance as a function of energy in a wide-band beam over a 1300 km baseline, 
\dune\ is highly sensitive to mass ordering, mixing parameters, and $\delta_{CP}$. Concurrently, \dune\ will test the internal
consistency of the three-flavor paradigm in a regime where matter effects and
interference among oscillation amplitudes are significant~\cite{bib26}.

In the context of an incomplete flavor structure, these measurements are not
only parameter determinations, but consistency tests. Deviations from the
expected oscillation patterns, even if small, may signal non-unitarity,
additional states, or modified propagation effects.

\subsection{The Near-Detector Complex as a Central Physics Instrument}

The \nd\ complex is central to the physics reach of \dune. In Phase I it comprises
the movable ND-LAr detector integrated with the downstream muon spectrometer (TMS),
together with the on-axis SAND detector~\cite{bib23,bib25}. By moving the ND-LAr+TMS system off axis
in the DUNE-PRISM configuration, different neutrino energy spectra are sampled; 
linear combinations of off-axis spectra can be used to produce data-driven far detector predictions 
with reduced interaction-model dependence~\cite{bib40}. The fixed on-axis SAND detector continuously 
monitors the beam and provides complementary measurements on argon and lighter 
targets, strengthening flux constraints and cross-section studies~\cite{bib41}.
Phase II's ND-GAr upgrade is specifically motivated by the need to push long-baseline 
systematics to the few-percent regime~\cite{bib23,bib45,bib46}.

The experimental advantage is not that neutrino scattering is free of complications,
but that the neutrino probe itself does not undergo QCD initial-state interactions before the
primary collision. In practice, precision still requires control of flux prediction,
nuclear effects, neutrino--nucleus interaction modeling, and detector response.
The \nd\ complex should therefore not be regarded as an auxiliary systematics tool,
but as the instrument that makes the long-baseline precision program robust while
also opening an independent program of rare-process and BSM searches~\cite{bib27}.

From the perspective developed in this work, the \nd\ can be viewed as a precision
probe of the neutrino-flavor sector. It enables measurements of:
\begin{itemize}
    \item flavor-dependent cross sections and their ratios,
    \item kinematic distributions sensitive to interaction dynamics,
    \item neutral-current and charged-current processes across channels,
    \item correlations among observables that test universality and consistency.
\end{itemize}

Such measurements are directly sensitive to the classes of deviations discussed
in the previous section, including non-standard interactions, non-unitarity,
and the possible presence of weakly coupled sectors.

\subsection{Near--far complementarity}

A defining feature of \dune\ is the complementarity between near and far detectors.
While the far detector probes neutrino propagation over long distances,
the \nd\ complex constrains the unoscillated beam and the neutrino-interaction
model near the source, especially for argon in the near-to-far transfer.
This combination provides the framework to separate and constrain:
\begin{itemize}
    \item propagation effects,
    \item interaction effects,
    \item detector systematics.
\end{itemize}

In particular, a consistent interpretation of oscillation data requires a precise
understanding of neutrino interactions. Conversely, deviations observed at the
\nd\ may acquire a different significance when combined with long-baseline
measurements. The joint analysis of near and far data therefore provides a
framework in which small, correlated deviations can be identified and
interpreted.

\dune, Hyper-Kamiokande, and JUNO should therefore be viewed as 
complementary rather than redundant precision programs. Their dominant 
systematics differ in character: Hyper-Kamiokande emphasizes enormous 
statistics in an off-axis narrow-band accelerator beam with a water-Cherenkov 
far detector, so its ultimate reach is closely tied to control of far-detector 
reconstruction and energy-scale systematics, whereas JUNO extracts oscillation 
information from reactor antineutrino spectra and is correspondingly sensitive to 
reactor-flux and spectral-shape uncertainties, reactor configuration, and exquisite 
energy-response control~\cite{bib42,bib43,bib44}. \dune\ contributes 
a distinct experimental balance---a 1300~km wide-band beam, liquid-argon imaging 
with strong particle-identification capability, and a controlled near-detector 
complex---which provides broad spectral leverage, detailed event characterization, 
and near--far consistency tests that are difficult to realize in the same way 
elsewhere~\cite{bib24,bib26}. Taken together, the three 
programs probe the same three-flavor framework with genuinely different 
experimental handles and failure modes.

% -------------------------------
\subsection{Illustrative Consistency-Test Strategy and Related Assumptions}

The practical objective, complementary to ongoing flavor-model-building efforts, is to convert DUNE's 
controlled redundancy in baseline, beam mode, detector subsystem, and final-state reconstruction into a 
staged closure test of the minimal three-flavor framework~\cite{bib15,bib16,bib23,bib24,bib25,bib45,bib46}. 
In the beam program, the combination of a wide-band beam, neutrino and antineutrino running, 
charged-current (CC) and neutral-current (NC) samples, a movable argon near-detector system, 
and an on-axis beam monitor gives DUNE access to an unusually broad fraction of the experimentally 
accessible neutrino-flavor space. The goal is therefore not only to measure standard oscillation 
parameters, but to test whether a single ND-constrained description closes simultaneously for 
all accessible channels and detector locations~\cite{bib23,bib25,bib40}.

\begin{table}[h]
\centering
\small
\setlength{\tabcolsep}{5pt}
\renewcommand{\arraystretch}{1.08}
\begin{tabular}{|p{0.465\linewidth}|p{0.465\linewidth}|}
\hline
\textbf{Near detector complex} & \textbf{Far detector} \\
\hline
\textbf{CC flavor-tagged samples:} high-statistics $\nu_\mu$ and $\nu_e$ inclusive and 
exclusive spectra in $\nu$ and $\bar\nu$ running; charge-sign information primarily from 
TMS in Phase~I and from ND-GAr in Phase~II; SAND provides on-axis monitoring and 
auxiliary interaction samples. \newline
\textbf{NC-sensitive samples:} inclusive NC rates, NC/CC ratios, $\pi^0/\gamma$-rich 
topologies, hadronic activity, and flavor-blind control samples. \newline
\textbf{Kinematics and topology:} lepton momentum and angle, hadronic visible energy, 
multiplicities, missing transverse momentum or imbalance observables, and PRISM 
dependence of reconstructed spectra. \newline
\textbf{Rare/BSM-sensitive channels:} trident-like signatures, long-lived-particle decays 
or scatters, displaced vertices, and rare electromagnetic topologies such as 
single-photon-like or missing-energy signatures where robust control samples can be defined. &
\textbf{CC flavor-tagged samples:} $\nu_e$ appearance and $\nu_\mu$ disappearance 
spectra in $\nu$ and $\bar\nu$ beam modes, together with $\nu_\tau$ appearance 
at higher energies. \newline
\textbf{NC-sensitive samples:} NC-enriched samples, flavor-blind depletion tests, and NC/CC 
cross-checks against ND-constrained expectations. \newline
\textbf{Kinematics and topology:} reconstructed lepton and hadronic energy, angular and 
topological residuals, and event-class migration patterns relative to the ND-constrained 
prediction. \newline
\textbf{Rare/BSM-sensitive channels:} anomalous appearance or disappearance, 
NC deficits, $\nu_\tau$ distortions, and rare FD channels that can be folded into the 
same global consistency program. \\
\hline
\end{tabular}
\caption{Schematic comparison of the dominant beam-program observables in the DUNE near and far detector systems.}
\label{tab:dune-nd-fd-observables}
\end{table}

\subsubsection{Comparison logic.}
At the conceptual level, the key comparison is not a naive ND/FD event ratio, but 
a hierarchy of constrained predictions. First, ND-LAr+TMS and SAND data determine the 
unoscillated beam composition, interaction model response, and time stability at the near site, 
supplemented by DUNE-PRISM off-axis sampling. Second, these constraints are propagated 
to the far site to predict the reconstructed FD spectra expected under the three-flavor hypothesis. 
A departure from the minimal framework would then appear as a failure of closure: 
energy-dependent residuals that cannot be absorbed by shared nuisance parameters, 
inconsistencies between appearance and disappearance channels, mismatches between 
neutrino and antineutrino running, anomalous NC/CC ratio shifts, or tension between topologically 
similar samples reconstructed in different detector subsystems~\cite{bib25,bib23,bib40,bib27}. 
In this sense, the most informative signal is not an isolated excess, but a coherent 
pattern of residuals across observables with different dependence on propagation, 
interaction modeling, and detector response.

\subsubsection{Conceptual joint fit.}
A natural way to organize this program is through a simultaneous ND+FD likelihood with shared nuisance parameters. A schematic implementation would be: 
\begin{equation}
\widehat{M}_{\rm FD}^{\rm ND} \equiv \mathcal{E}_{\rm ND\to FD} \left( D_{\rm ND}; \boldsymbol{\theta}, \boldsymbol{\xi}, \boldsymbol{\eta}_{\rm sh}, \boldsymbol{\eta}_{\rm ND} \right)
\end{equation}

\begin{equation}
\begin{split} 
\chi^2_{\rm joint}(\boldsymbol{\theta}, \boldsymbol{\xi}, \boldsymbol{\eta}_{\rm sh}, \boldsymbol{\eta}_{\rm ND}, \boldsymbol{\eta}_{\rm FD}) = \, & \chi^2_{\rm ND} \left[ D_{\rm ND} \middle| M_{\rm ND}(\boldsymbol{\theta}, \boldsymbol{\xi}, \boldsymbol{\eta}_{\rm sh}, \boldsymbol{\eta}_{\rm ND}) \right] \\ 
& + \chi^2_{\rm FD} \left( D_{\rm FD} \middle| \widehat{M}_{\rm FD}^{\rm ND},\boldsymbol{\eta}_{\rm FD} \right) \\ 
& + \chi^2_{\rm prior}(\boldsymbol{\eta}_{\rm sh}, \boldsymbol{\eta}_{\rm ND}, \boldsymbol{\eta}_{\rm FD}) .
\end{split}
\end{equation}

Here $\boldsymbol{\theta}$ denotes the standard oscillation parameters, $\boldsymbol{\xi}$ a 
benchmark set of BSM parameters (for example non-unitarity, NSI, or sterile-neutrino admixtures), 
$\boldsymbol{\eta}_{\rm sh}$ the nuisance parameters shared between ND and FD (flux, interaction, 
and common calibration terms), and $\boldsymbol{\eta}_{\rm ND}$ and $\boldsymbol{\eta}_{\rm FD}$ 
detector-specific response terms. In practice, $\chi^2_{\rm ND}$ should be understood as a sum over 
ND subsystems and PRISM positions, while $\chi^2_{\rm prior}$ encodes external constraints and 
calibration information. A sensible implementation of this strategy is then: first fit the standard three-flavor closure 
hypothesis; then perform blinded closure and injection tests on pseudo-data; and only then compare 
benchmark BSM hypotheses against the same constrained data model~\cite{bib40,bib27,bib23}.

\subsubsection{Assumptions.}
This strategy rests on a small set of explicit assumptions. The near-to-far flux relation and beamline 
geometry must be controlled well enough that PRISM-weighted ND spectra can be meaningfully 
propagated to the FD. 
Detailed sterile-neutrino benchmark studies already show that DUNE’s combined ND+FD exclusion 
reach depends non-negligibly on modeling the extended neutrino source and the resulting baseline 
smearing, and that this effect is not washed out by generic uncorrelated shape systematics.~\cite{bib48}.
Acceptance and efficiency differences between ND subsystems and the FD 
must be modeled or calibrated rather than assumed to cancel. Nuclear-model uncertainties, including 
final-state interactions and missing energy, must enter explicitly through nuisance parameters and 
validation samples, not be silently reabsorbed into apparent flavor anomalies. Detector-response 
differences between ND-LAr, TMS/SAND, and the FD must remain factorized enough that shared 
and detector-specific covariance terms are interpretable. What is intentionally left unspecified here 
is the detailed binning, the exact regularization of PRISM weights, and the full covariance 
parameterization of nuclear and detector systematics; these belong in a dedicated technical 
analysis note rather than in the present perspective~\cite{bib25,bib45,bib46,bib40}.

\subsubsection{Limitations and expectations.}
This program should be understood as a structured self-consistency test, not as a claim that DUNE 
alone can solve the flavor problem. Phase~I already enables meaningful closure and residual-based 
tests, but no single universal public systematic floor exists for the broader class of searches advocated 
here. Phase~II, with ND-GAr, broader acceptance, lower thresholds, and stronger charge-sign capability, 
is the natural regime in which DUNE aims to push long-baseline systematic control to the few-percent 
level required by its ultimate precision goals. For that reason no single numerical discovery reach is 
quoted here: the relevant sensitivity is channel- and hypothesis-dependent, and published projections 
remain benchmark-specific rather than universal~\cite{bib23,bib45,bib46,bib40,bib47,bib27}.
The fit equations and comparison logic described in this section are illustrative and schematic, 
not a proposal for the official DUNE analysis framework.

%-------------------------------

\subsection{Toward an expanded physics role}

Within this perspective, the role of \dune\ extends beyond the measurement of
oscillation parameters. It becomes a program capable of testing the self-consistency
and possible limits of the present three-flavor framework, while searching for
small departures that may point to deeper neutrino-flavor structure. The \nd, in
particular, should not be regarded solely as an auxiliary instrument, but as a
central component of this broader physics program~\cite{bib27}.

By enabling precision measurements across multiple observables and channels,
\dune\ offers the opportunity to test the internal consistency of the
neutrino-flavor sector with exceptional sensitivity. In doing so, it may help
define more sharply the empirical domain within which any deeper theory of
flavor must operate.

% -------------------------------------------------
\section{Conclusions} 
% -------------------------------------------------

The replication of fermion families, the hierarchy of masses,
and the special role of neutrinos all suggest that flavor may point
to a deeper organizing principle beyond the \sm.
Despite its empirical success, the flavor sector remains structurally incomplete:
it is fully operational, yet lacks a unifying framework comparable to that of
the gauge interactions.

This work does not propose a specific mechanism.
Rather, it emphasizes that flavor should be treated as a sector 
whose underlying organization is not yet understood,
and that experimental programs such as \dune\ can play a central role
in probing its limits.

The neutrino sector is where that incompleteness becomes experimentally sharp.
\dune\ will not solve the flavor problem in full, but it will provide one of the most powerful
tests of the three-flavor framework through its 1300 km wide-band long-baseline
measurement and its Phase I near-detector complex---ND-LAr+TMS, SAND,
and DUNE-PRISM. Together, these systems are designed to reduce interaction-model
dependence, strengthen data-driven near-to-far predictions, and search for correlated
departures from standard oscillation and interaction expectations, including non-unitarity,
non-standard interactions, additional neutrino states, and beam-produced BSM signatures.
In that sense, \dune\ is not only a precision measurement program; it is an experiment
that will define more sharply the empirical domain within which any deeper theory of flavor must operate.

If precision neutrino data increasingly delimit the viable low-energy domain of flavor models without revealing 
a unique organizing principle, they will also sharpen the case for future precision facilities.

\section{Acknowledgments}

This manuscript has been authored by Fermi Forward Discovery Group 
under Contract No. 89243024CSC000002 with the U.S.
Department of Energy, Office of Science, Office of High Energy Physics.

The author acknowledges the use of OpenAI's ChatGPT and Ask Sage as tools for language 
refinement, editorial support, and conceptual organization during the preparation of this manuscript. 
All scientific content, physics arguments, and conclusions are the author's original contributions.

The author also thanks colleagues and collaborators for helpful discussions and
for providing a scientific environment that encourages broad theoretical
exploration.

% -------------------------------------------------
% Bibliography placeholder
% -------------------------------------------------

\bibliography{DUNE_Flavor_V2}

@article{bib1,
  title={Broken Symmetries and the Masses of Gauge Bosons},
  author={Higgs, Peter W.},
  journal={Phys. Rev. Lett.},
  volume = {13},
  issue = {16},
  pages = {508--509},
  year = {1964},
  month = {Oct},
  publisher = {American Physical Society},
  doi = {10.1103/PhysRevLett.13.508},
  url = {https://link.aps.org/doi/10.1103/PhysRevLett.13.508}
}

@article{bib2,
  title = {Broken Symmetry and the Mass of Gauge Vector Mesons},
  author = {Englert, F. and Brout, R.},
  journal = {Phys. Rev. Lett.},
  volume = {13},
  issue = {9},
  pages = {321--323},
  numpages = {0},
  year = {1964},
  month = {Aug},
  publisher = {American Physical Society},
  doi = {10.1103/PhysRevLett.13.321},
  url = {https://link.aps.org/doi/10.1103/PhysRevLett.13.321}
}

@article{bib2_1,
  title = {Global Conservation Laws and Massless Particles},
  author = {Guralnik, G. S. and Hagen, C. R. and Kibble, T. W. B.},
  journal = {Phys. Rev. Lett.},
  volume = {13},
  issue = {20},
  pages = {585--587},
  numpages = {0},
  year = {1964},
  month = {Nov},
  publisher = {American Physical Society},
  doi = {10.1103/PhysRevLett.13.585},
  url = {https://link.aps.org/doi/10.1103/PhysRevLett.13.585}
}

@article{bib3,
title = {Observation of a new particle in the search for the Standard Model Higgs boson with the ATLAS detector at the LHC},
author = {G. Aad and others (ATLAS Collaboration)},
journal = {Physics Letters B},
volume = {716},
number = {1},
pages = {1-29},
year = {2012},
issn = {0370-2693},
doi = {https://doi.org/10.1016/j.physletb.2012.08.020},
url = {https://www.sciencedirect.com/science/article/pii/S037026931200857X}
}

@article{bib4,
title = {Observation of a new boson at a mass of 125 GeV with the CMS experiment at the LHC},
author = {S. Chatrchyan and others (CMS Collaboration)},
journal = {Physics Letters B},
volume = {716},
number = {1},
pages = {30-61},
year = {2012},
issn = {0370-2693},
doi = {https://doi.org/10.1016/j.physletb.2012.08.021},
url = {https://www.sciencedirect.com/science/article/pii/S0370269312008581}
}

@article{bib5,
  title = {Varieties of baryon and lepton nonconservation},
  author = {Weinberg, Steven},
  journal = {Phys. Rev. D},
  volume = {22},
  issue = {7},
  pages = {1694--1700},
  numpages = {0},
  year = {1980},
  month = {Oct},
  publisher = {American Physical Society},
  doi = {10.1103/PhysRevD.22.1694},
  url = {https://link.aps.org/doi/10.1103/PhysRevD.22.1694}
}

@article{bib6,
  title     = {Mesonium and Antimesonium},
  author    = {Pontecorvo, B.},
  journal   = {Sov. Phys. JETP},
  volume    = {6},
  pages     = {429--431},
  year      = {1958}
}

@article{bib7,
    title = {Remarks on the Unified Model of Elementary Particles},
     author = {Maki, Ziro and Nakagawa, Masami and Sakata, Shoichi},
   journal = {Progress of Theoretical Physics},
    volume = {28},
    number = {5},
    pages = {870-880},
    year = {1962},
    month = {11},
    doi = {10.1143/PTP.28.870},
    url = {https://doi.org/10.1143/PTP.28.870},
    eprint = {https://academic.oup.com/ptp/article-pdf/28/5/870/5258750/28-5-870.pdf}
}

@article{bib8,
    title = {NuFit-6.0: updated global analysis of three-flavor neutrino oscillations},
     author = {Esteban, Ivan and Gonzalez-Garcia, M. C. and Maltoni, Michele and Martinez-Soler, Ivan and Pinheiro, João Paulo and Schwetz, Thomas},
   journal = {Journal of High Energy Physics},
    volume = {2024},
    number = {12},
    pages = {216},
    year = {2024},
    doi = {10.1007/JHEP12(2024)216},
    url = {https://doi.org/10.1007/JHEP12(2024)216}
}

@article{bib9,
  title = {Evidence for Oscillation of Atmospheric Neutrinos},
  author = {Fukuda, Y. and Hayakawa, T. and Ichihara, E. and Inoue, K. and Ishihara, K. and Ishino, H. and Itow, Y. and Kajita, T. and Kameda, J. and Kasuga, S. and Kobayashi, K. and Kobayashi, Y. and Koshio, Y. and Miura, M. and Nakahata, M. and Nakayama, S. and Okada, A. and Okumura, K. and others},
  collaboration = {Super-Kamiokande Collaboration},
  journal = {Phys. Rev. Lett.},
  volume = {81},
  issue = {8},
  pages = {1562--1567},
  numpages = {0},
  year = {1998},
  month = {Aug},
  publisher = {American Physical Society},
  doi = {10.1103/PhysRevLett.81.1562},
  url = {https://link.aps.org/doi/10.1103/PhysRevLett.81.1562}
}

@article{bib10,
  title = {Direct Evidence for Neutrino Flavor Transformation from Neutral-Current Interactions in the Sudbury Neutrino Observatory},
  author = {Ahmad, Q. R. and Allen, R. C. and Andersen, T. C. and D.Anglin, J. and Barton, J. C. and Beier, E. W. and Bercovitch, M. and Bigu, J. and Biller, S. D. and Black, R. A. and Blevis, I. and Boardman, R. J. and Boger, J. and Bonvin, E. and Boulay, M. G. and Bowler, M. G. and Bowles, T. J. and Brice, S. J. and others},
  collaboration = {SNO Collaboration},
  journal = {Phys. Rev. Lett.},
  volume = {89},
  issue = {1},
  pages = {011301},
  numpages = {6},
  year = {2002},
  month = {Jun},
  publisher = {American Physical Society},
  doi = {10.1103/PhysRevLett.89.011301},
  url = {https://link.aps.org/doi/10.1103/PhysRevLett.89.011301}
}

@article{bib11,
  title = {First Results from KamLAND: Evidence for Reactor Antineutrino Disappearance},
  author = {Eguchi, K. and Enomoto, S. and Furuno, K. and Goldman, J. and Hanada, H. and Ikeda, H. and Ikeda, K. and Inoue, K. and Ishihara, K. and Itoh, W. and Iwamoto, T. and Kawaguchi, T. and Kawashima, T. and Kinoshita, H. and Kishimoto, Y. and Koga, M. and Koseki, Y. and Maeda, T. and others},
  collaboration = {KamLAND Collaboration},
  journal = {Phys. Rev. Lett.},
  volume = {90},
  issue = {2},
  pages = {021802},
  numpages = {6},
  year = {2003},
  month = {Jan},
  publisher = {American Physical Society},
  doi = {10.1103/PhysRevLett.90.021802},
  url = {https://link.aps.org/doi/10.1103/PhysRevLett.90.021802}
}

@article{bib12,
title = {{$\mu \rightarrow e \gamma$ at a rate of one out of $10^9$ muon decays?}},
author = {Peter Minkowski},
journal = {Physics Letters B},
volume = {67},
number = {4},
pages = {421-428},
year = {1977},
issn = {0370-2693},
doi = {https://doi.org/10.1016/0370-2693(77)90435-X},
url = {https://www.sciencedirect.com/science/article/pii/037026937790435X}
}

@article{bib13,
  title = {{Neutrino masses, mixings, and oscillations in $\mathrm{SU}(2) \times \mathrm{U}(1)$ models of electroweak interactions}},
  author = {Cheng, T. P. and Li, Ling-Fong},
  journal = {Phys. Rev. D},
  volume = {22},
  issue = {11},
  pages = {2860--2868},
  numpages = {0},
  year = {1980},
  month = {Dec},
  publisher = {American Physical Society},
  doi = {10.1103/PhysRevD.22.2860},
  url = {https://link.aps.org/doi/10.1103/PhysRevD.22.2860}
}

@article{bib14,
  title = {Neutrino Mass and Spontaneous Parity Nonconservation},
  author = {Mohapatra, Rabindra N. and Senjanovi\'{c}, Goran},
  journal = {Phys. Rev. Lett.},
  volume = {44},
  issue = {14},
  pages = {912--915},
  numpages = {0},
  year = {1980},
  month = {Apr},
  publisher = {American Physical Society},
  doi = {10.1103/PhysRevLett.44.912},
  url = {https://link.aps.org/doi/10.1103/PhysRevLett.44.912}
}

@article{bib15,
  title = {Lepton flavor symmetries},
  author = {Feruglio, Ferruccio and Romanino, Andrea},
  journal = {Rev. Mod. Phys.},
  volume = {93},
  issue = {1},
  pages = {015007},
  numpages = {50},
  year = {2021},
  month = {Mar},
  publisher = {American Physical Society},
  doi = {10.1103/RevModPhys.93.015007},
  url = {https://link.aps.org/doi/10.1103/RevModPhys.93.015007}
}

@article{bib16,
title = {Phenomenology of lepton masses and mixing with discrete flavor symmetries},
author = {Garv Chauhan and P.S. Bhupal Dev and Ievgen Dubovyk and Bartosz Dziewit and Wojciech Flieger and Krzysztof Grzanka and Janusz Gluza and Biswajit Karmakar and Szymon Zieba},
journal = {Progress in Particle and Nuclear Physics},
volume = {138},
pages = {104126},
year = {2024},
issn = {0146-6410},
doi = {https://doi.org/10.1016/j.ppnp.2024.104126},
url = {https://www.sciencedirect.com/science/article/pii/S0146641024000309}
}

@article{bib17,
  title = {Neutrino oscillations in matter},
  author = {Wolfenstein, L.},
  journal = {Phys. Rev. D},
  volume = {17},
  issue = {9},
  pages = {2369--2374},
  numpages = {0},
  year = {1978},
  month = {May},
  publisher = {American Physical Society},
  doi = {10.1103/PhysRevD.17.2369},
  url = {https://link.aps.org/doi/10.1103/PhysRevD.17.2369}
}

@article{bib18,
title = {On the MSW effect with massless neutrinos and no mixing in the vacuum},
author = {M.M. Guzzo and A. Masiero and S.T. Petcov},
journal = {Physics Letters B},
volume = {260},
number = {1},
pages = {154-160},
year = {1991},
issn = {0370-2693},
doi = {https://doi.org/10.1016/0370-2693(91)90984-X},
url = {https://www.sciencedirect.com/science/article/pii/037026939190984X}
}

@article{bib19,
title = {Status of non-standard neutrino interactions},
author = {Ohlsson, Tommy},
journal = {Reports on Progress in Physics},
year = {2013},
month = {mar},
publisher = {IOP Publishing},
volume = {76},
number = {4},
pages = {044201},
doi = {10.1088/0034-4885/76/4/044201},
url = {https://doi.org/10.1088/0034-4885/76/4/044201}
}

@ARTICLE{bib20,
title={Neutrino Oscillations and Non-standard Interactions},         
author={Farzan, Yasaman  and Tórtola, Mariam },           
journal={Frontiers in Physics},          
volume={6},  
year={2018},  
url={https://www.frontiersin.org/journals/physics/articles/10.3389/fphy.2018.00010},  
doi={10.3389/fphy.2018.00010},  
issn={2296-424X}
}

@article{bib21,
title={Light Sterile Neutrinos: A White Paper}, 
author={K. N. Abazajian and M. A. Acero and S. K. Agarwalla and A. A. Aguilar-Arevalo and C. H. Albright and S. Antusch and C. A. Arguelles and A. B. Balantekin and G. Barenboim and V. Barger and P. Bernardini and F. Bezrukov and O. E. Bjaelde and S. A. Bogacz and N. S. Bowden and A. Boyarsky and others},
year={2012},
eprint={1204.5379},
archivePrefix={arXiv},
primaryClass={hep-ph},
url={https://arxiv.org/abs/1204.5379}
}

@article{bib22,
title = {Where are we with light sterile neutrinos?},
author = {A. Diaz and C.A. Argüelles and G.H. Collin and J.M. Conrad and M.H. Shaevitz},
journal = {Physics Reports},
volume = {884},
pages = {1-59},
year = {2020},
note = {Where are we with light sterile neutrinos?},
issn = {0370-1573},
doi = {https://doi.org/10.1016/j.physrep.2020.08.005},
url = {https://www.sciencedirect.com/science/article/pii/S0370157320302921}
}

@article{bib23,
title={The DUNE Science Program}, 
author={A. Abed Abud and R. Acciarri and M. A. Acero and M. R. Adames and G. Adamov and M. Adamowski and D. Adams and M. Adinolfi and C. Adriano and A. Aduszkiewicz and J. Aguilar and F. Akbar and F. Alemanno and N. S. Alex and K. Allison and M. Alrashed and A. Alton and others},
collaboration = {DUNE Collaboration},
year={2025},
eprint={2503.23291},
archivePrefix={arXiv},
primaryClass={hep-ex},
url={https://arxiv.org/abs/2503.23291}, 
}

@article{bib24,
  title={Deep Underground Neutrino Experiment (DUNE), Far Detector Technical Design Report, Volume I. Introduction to DUNE},
  author={Abi, B. and Acciarri, R. and Acero, M.A. and others},
  collaboration = {DUNE Collaboration},
  journal={Journal of Instrumentation},
  volume={15},
  number={08},
  pages={T08008},
  year={2020},
  month={aug}
}

@Article{bib25,
title= {Deep Underground Neutrino Experiment (DUNE) Near Detector Conceptual Design Report},
author={Abud, A. Abed and Abi, B. and Acciarri, R. and Acero, M. A. and others},
collaboration = {DUNE Collaboration},
journal={Instruments},
volume={5},
year={2021},
number={4},
article-number={31},
issn={2410-390X},
doi={10.3390/instruments5040031},
url={https://www.mdpi.com/2410-390X/5/4/31}
}

@Article{bib26,
title= {Long-baseline neutrino oscillation physics potential of the DUNE experiment},
author={Abi, B. and Acciarri, R. and Acero, M. A. and Adamov, G. and others},
collaboration = {DUNE Collaboration},
journal={The European Physical Journal C},
volume={80},
year={2020},
number={10},
pages={978},
isbn={1434-6052},
doi={10.1140/epjc/s10052-020-08456-z},
url={https://doi.org/10.1140/epjc/s10052-020-08456-z}
}

@Article{bib27,
title= {Prospects for beyond the Standard Model physics searches at the Deep Underground Neutrino Experiment},
author={Abi, B. and Acciarri, R. and Acero, M. A. and Adamov, G. and Adams, D. and others},
collaboration = {DUNE Collaboration},
journal={The European Physical Journal C},
volume={81},
year={2021},
number={4},
pages={322},
isbn={1434-6052},
doi={10.1140/epjc/s10052-021-09007-w},
url={https://doi.org/10.1140/epjc/s10052-021-09007-w}
}

@article{bib28,
  title = {Evidence for ${\overline{\ensuremath{\nu}}}_{\ensuremath{\mu}}\ensuremath{\rightarrow}{\overline{\ensuremath{\nu}}}_{\mathit{e}}$ Oscillations from the LSND Experiment at the Los Alamos Meson Physics Facility},
  author = {Athanassopoulos, C. and Auerbach, L. B. and Burman, R. L. and others},
  collaboration = {LSND Collaboration},
  journal = {Phys. Rev. Lett.},
  volume = {77},
  issue = {15},
  pages = {3082--3085},
  numpages = {0},
  year = {1996},
  month = {Oct},
  publisher = {American Physical Society},
  doi = {10.1103/PhysRevLett.77.3082},
  url = {https://link.aps.org/doi/10.1103/PhysRevLett.77.3082}
}

@article{bib29,
  title = {Significant Excess of Electronlike Events in the MiniBooNE Short-Baseline Neutrino Experiment},
  author = {Aguilar-Arevalo, A. A. and Brown, B. C. and Bugel, L. and Cheng, G. and Conrad, J. M. and Cooper, R. L. and Dharmapalan, R. and others},
  collaboration = {MiniBooNE Collaboration},
  journal = {Phys. Rev. Lett.},
  volume = {121},
  issue = {22},
  pages = {221801},
  numpages = {7},
  year = {2018},
  month = {Nov},
  publisher = {American Physical Society},
  doi = {10.1103/PhysRevLett.121.221801},
  url = {https://link.aps.org/doi/10.1103/PhysRevLett.121.221801}
}

@article{bib30,
  title = {Reactor antineutrino anomaly},
  author = {Mention, G. and Fechner, M. and Lasserre, Th. and Mueller, Th. A. and Lhuillier, D. and Cribier, M. and Letourneau, A.},
  journal = {Phys. Rev. D},
  volume = {83},
  issue = {7},
  pages = {073006},
  numpages = {20},
  year = {2011},
  month = {Apr},
  publisher = {American Physical Society},
  doi = {10.1103/PhysRevD.83.073006},
  url = {https://link.aps.org/doi/10.1103/PhysRevD.83.073006}
}

@article{bib31,
  title = {Statistical significance of the gallium anomaly},
  author = {Giunti, Carlo and Laveder, Marco},
  journal = {Phys. Rev. C},
  volume = {83},
  issue = {6},
  pages = {065504},
  numpages = {5},
  year = {2011},
  month = {Jun},
  publisher = {American Physical Society},
  doi = {10.1103/PhysRevC.83.065504},
  url = {https://link.aps.org/doi/10.1103/PhysRevC.83.065504}
}

@article{bib31_1,
  title = {Search for light sterile neutrinos with two neutrino beams at MicroBooNE},
  author = {Abratenko, P. and Aldana, D. Andrade and Arellano, L. and Asaadi, J. and Ashkenazi, A. and others},
  collaboration = {MicroBooNE Collaboration},
  journal = {Nature},
  volume = {648},
  issue = {8092},
  pages = {64--69},
  year = {2025},
  isbn={1476-4687},
  doi = {10.1038/s41586-025-09757-7},
  url = {https://doi.org/10.1038/s41586-025-09757-7}
}

@article{bib32,
  title = {STEREO neutrino spectrum of 235U fission rejects sterile neutrino hypothesis},
  author = {Almazán, H. and Bernard, L. and Blanchet, A. and Bonhomme, A. and Buck, C. and Chalil, A. and others},
  collaboration = {STEREO Collaboration},
  journal = {Nature},
  volume = {613},
  issue = {7943},
  pages = {257--261},
  year = {2023},
  isbn={1476-4687},
  doi = {10.1038/s41586-022-05568-2},
  url = {https://doi.org/10.1038/s41586-022-05568-2}
}

@article{bib33,
title = {Reactor antineutrino flux and anomaly},
author = {Chao Zhang and Xin Qian and Muriel Fallot},
journal = {Progress in Particle and Nuclear Physics},
volume = {136},
pages = {104106},
year = {2024},
issn = {0146-6410},
doi = {https://doi.org/10.1016/j.ppnp.2024.104106},
url = {https://www.sciencedirect.com/science/article/pii/S0146641024000103}
}

@article{bib34,
  title = {Sterile-neutrino search based on 259 days of KATRIN data},
  author = {Acharya, H. and Aker, M. and Batzler, D. and Beglarian, A. and Beisenkötter, J. and Biassoni, M. and Bieringer, B. and others},
  collaboration = {KATRIN Collaboration},
  year = {2025},
  journal = {Nature},
  volume = {648},
  issue = {8092},
  pages = {70--75},
  isbn={1476-4687},
  doi = {10.1038/s41586-025-09739-9},
  url = {https://doi.org/10.1038/s41586-025-09739-9}
}

@article{bib35,
  title = {{Gallium Anomaly: critical view from the global picture of $\nu_e$ and $\bar{\nu}_e$ disappearance}},
  author = {Giunti, C. and Li, Y. F. and Ternes, C. A. and Tyagi, O. and Xin, Z.},
  journal = {Journal of High Energy Physics},
  year = {2022},
  volume = {2022},
  issue = {10},
  pages = {164},
  isbn={1029-8479},
  doi = {10.1007/JHEP10(2022)164},
  url = {https://doi.org/10.1007/JHEP10(2022)164}
}

@article{bib36,
title = {The gallium anomaly},
author = {S.R. Elliott and V.N. Gavrin and W.C. Haxton},
journal = {Progress in Particle and Nuclear Physics},
volume = {134},
pages = {104082},
year = {2024},
issn = {0146-6410},
doi = {https://doi.org/10.1016/j.ppnp.2023.104082},
url = {https://www.sciencedirect.com/science/article/pii/S0146641023000637}
}

@article{bib37,
  title = {The science and technology of liquid argon detectors},
  author = {Bonivento, W. M. and Terranova, F.},
  journal = {Rev. Mod. Phys.},
  volume = {96},
  issue = {4},
  pages = {045001},
  numpages = {40},
  year = {2024},
  month = {Oct},
  publisher = {American Physical Society},
  doi = {10.1103/RevModPhys.96.045001},
  url = {https://link.aps.org/doi/10.1103/RevModPhys.96.045001}
}

@article{bib38,
  title = {Search for an Excess of Electron Neutrino Interactions in MicroBooNE Using Multiple Final-State Topologies},
  author = {Abratenko, P. and An, R. and Anthony, J. and Arellano, L. and Asaadi, J. and Ashkenazi, A. and others},
  collaboration = {MicroBooNE Collaboration},
  journal = {Phys. Rev. Lett.},
  volume = {128},
  issue = {24},
  pages = {241801},
  numpages = {9},
  year = {2022},
  month = {Jun},
  publisher = {American Physical Society},
  doi = {10.1103/PhysRevLett.128.241801},
  url = {https://link.aps.org/doi/10.1103/PhysRevLett.128.241801}
}

@article{bib39,
  title = {Search for Neutrino-Induced Neutral-Current $\mathrm{\ensuremath{\Delta}}$ Radiative Decay in MicroBooNE and a First Test of the MiniBooNE Low Energy Excess under a Single-Photon Hypothesis},
  author = {Abratenko, P. and An, R. and Anthony, J. and Arellano, L. and Asaadi, J. and Ashkenazi, A. and others},
  collaboration = {MicroBooNE Collaboration},
  journal = {Phys. Rev. Lett.},
  volume = {128},
  issue = {11},
  pages = {111801},
  numpages = {8},
  year = {2022},
  month = {Mar},
  publisher = {American Physical Society},
  doi = {10.1103/PhysRevLett.128.111801},
  url = {https://link.aps.org/doi/10.1103/PhysRevLett.128.111801}
}

@article{bib40,
title={DUNE-PRISM: Reducing neutrino interaction model dependence with a movable neutrino detector}, 
author={Ciaran Hasnip},
collaboration = {DUNE Collaboration},
year={2025},
eprint={2501.14811},
archivePrefix={arXiv},
primaryClass={hep-ex},
url={https://arxiv.org/abs/2501.14811}, 
}

@article{bib41,
title = {The SAND detector of the DUNE experiment},
author = {Nicolò Tosi},
collaboration = {DUNE Collaboration},
journal = {Nuclear Instruments and Methods in Physics Research Section A: Accelerators, Spectrometers, Detectors and Associated Equipment},
volume = {1080},
pages = {170727},
year = {2025},
issn = {0168-9002},
doi = {https://doi.org/10.1016/j.nima.2025.170727},
url = {https://www.sciencedirect.com/science/article/pii/S0168900225005285}
}

@article{bib42,
title={Hyper-Kamiokande Design Report}, 
author={K. Abe and Ke. Abe and H. Aihara and others},
collaboration = {Hyper-Kamiokande Proto-Collaboration},
year={2018},
eprint={1805.04163},
archivePrefix={arXiv},
primaryClass={physics.ins-det},
url={https://arxiv.org/abs/1805.04163}
}

@article{bib43,
  title = {Sensitivity of the Hyper-Kamiokande experiment to neutrino oscillation parameters using accelerator neutrinos},
  author = {Abe, K. and Afif, M. T. and Ahl Laamara, R. and Aihara, H. and Ajmi, A. and Akutsu, R. and others},
  collaboration = {Hyper-Kamiokande Collaboration},
  journal = {The European Physical Journal C},
  volume = {86},
  issue = {2},
  pages = {170},
  year = {2026},
  month = {Feb},
  publisher = {Springer Berlin Heidelberg},
  doi = {10.1140/epjc/s10052-025-14938-9},
  url = {https://doi.org/10.1140/epjc/s10052-025-14938-9}
}

@article{bib44,
title = {JUNO physics and detector},
author = {JUNO Collaboration},
journal = {Progress in Particle and Nuclear Physics},
volume = {123},
pages = {103927},
year = {2022},
issn = {0146-6410},
publisher={Elsevier BV},
doi = {https://doi.org/10.1016/j.ppnp.2021.103927},
url = {https://www.sciencedirect.com/science/article/pii/S0146641021000880}
}

@article{bib45,
title = {DUNE Phase II: scientific opportunities, detector concepts, technological solutions},
author = {Abed Abud, A. and Abi, B. and Acciarri, R. and Acero, M.A. and Adames, M.R. and Adamov, G. and Adamowski, M. and Adams, D. and others},
collaboration = {DUNE Collaboration},
journal = {Journal of Instrumentation},
year = {2024},
month = {dec},
publisher = {IOP Publishing},
volume = {19},
number = {12},
pages = {P12005},
doi = {10.1088/1748-0221/19/12/P12005},
url = {https://doi.org/10.1088/1748-0221/19/12/P12005}
}

@article{bib46,
title={The DUNE Phase II Detectors}, 
author={A. Abed Abud and R. Acciarri and M. A. Acero and M. R. Adames and G. Adamov and M. Adamowski and D. Adams and others},
collaboration = {DUNE Collaboration},
year={2025},
eprint={2503.23293},
archivePrefix={arXiv},
primaryClass={physics.ins-det},
url={https://arxiv.org/abs/2503.23293}
}

@article{bib47,
  title = {Physics with beam tau-neutrino appearance at DUNE},
  author = {de Gouv\^ea, Andr\'e and Kelly, Kevin J. and Stenico, G. V. and Pasquini, Pedro},
  journal = {Phys. Rev. D},
  volume = {100},
  issue = {1},
  pages = {016004},
  numpages = {18},
  year = {2019},
  month = {Jul},
  publisher = {American Physical Society},
  doi = {10.1103/PhysRevD.100.016004},
  url = {https://link.aps.org/doi/10.1103/PhysRevD.100.016004}
}

@article{bib48,
  title = {Baseline and other effects for a sterile neutrino at DUNE},
  author = {Penedo, J. T. and Pulido, Jo\~ao},
  journal = {Phys. Rev. D},
  volume = {107},
  issue = {7},
  pages = {075026},
  numpages = {16},
  year = {2023},
  month = {Apr},
  publisher = {American Physical Society},
  doi = {10.1103/PhysRevD.107.075026},
  url = {https://link.aps.org/doi/10.1103/PhysRevD.107.075026}
}

\end{document}